\documentclass[reprint,showpacs,amsmath,amssymb,floatfix,superscriptaddress,aps,prb]{revtex4-1}

\usepackage{graphicx}
\usepackage{float}
\usepackage{color}
\usepackage{dcolumn}
\usepackage{bm}

\begin{document}

\title{Antiferromagnetic structure and electronic properties
of BaCr$_2$As$_2$ and BaCrFeAs$_2$}

\author{Kai A.\ Filsinger}
\author{Walter Schnelle}
\author{Peter Adler}
\author{Gerhard H.\ Fecher}
\affiliation{Max Planck Institute for Chemical Physics of Solids,
N\"othnitzer Stra{\ss}e 40, 01187 Dresden, Germany}
\author{Manfred Reehuis}
\author{Andreas Hoser}
\author{Jens-Uwe Hoffmann}
\affiliation{Helmholtz-Zentrum Berlin f\"ur Materialien und Energie,
14109 Berlin, Germany}
\author{Peter Werner}
\affiliation{Max Planck Institute of Microstructure Physics,
Weinberg 2, 06120 Halle (Saale), Germany}
\author{Martha Greenblatt}
\affiliation{Department of Chemistry and Chemical Biology, Rutgers University,
Piscataway, NJ, USA}
\author{Claudia Felser}
\affiliation{Max Planck Institute for Chemical Physics of Solids,
N\"othnitzer Stra{\ss}e 40, 01187 Dresden, Germany}

%%%%%%%%%%%%%%%%%%%%%%%%%%%%%%%%%%%%%%%%%%%%%%%%%%%%%%%%%%%%%%%%%%%%%%%%%%%%%%%

\date{\today}

\begin{abstract}
The chromium arsenides BaCr$_2$As$_2$ and BaCrFeAs$_2$ with ThCr$_2$Si$_2$
type structure (space group $I4/mmm$; also adopted by `122' iron arsenide
superconductors) have been suggested as mother compounds for possible new
superconductors. DFT-based calculations of the electronic structure evidence
metallic antiferromagnetic ground states for both compounds. By powder neutron
diffraction we confirm for BaCr$_2$As$_2$ a robust ordering in the
antiferromagnetic $G$-type structure at $T_\mathrm{N} = 580$\,K with
$\mu_\mathrm{Cr} = 1.9\,\mu_\mathrm{B}$ at $T = 2$\,K. Anomalies in the
lattice parameters point to magneto-structural coupling effects. In
BaCrFeAs$_2$ the Cr and Fe atoms randomly occupy the transition-metal site and
$G$-type order is found below 265\,K with $\mu_\mathrm{Cr/Fe} =
1.1\,\mu_\mathrm{B}$. $^{57}$Fe M\"ossbauer spectroscopy demonstrates that
only a small ordered moment is associated with the Fe atoms, in agreement with
electronic structure calculations with $\mu_\mathrm{Fe} \sim 0$. The
temperature dependence of the hyperfine field does not follow that of the
total moments. Both compounds are metallic but show large enhancements of the
linear specific heat coefficient $\gamma$ with respect to the band structure
values. The metallic state and the electrical transport in BaCrFeAs$_2$ is
dominated by the atomic disorder of Cr and Fe and partial magnetic disorder of
Fe. Our results indicate that N\'eel-type order is unfavorable for the Fe
moments and thus it is destabilized with increasing iron content.
\end{abstract}

\maketitle

%%%%%%%%%%%%%%%%%%%%%%%%%%%%%%%%%%%%%%%%%%%%%%%%%%%%%%%%%%%%%%%%%%%%%%%%%%%%%%%

\section{Introduction}
\label{sec:intro}

The discovery of high-$T_\mathrm{c}$ superconductivity in iron-based pnictides
and chalcogenides has initiated intense efforts to investigate and
theoretically model the magnetic states adopted by the non-superconducting
relatives of these compounds [\onlinecite{BAC16,DAI15,SiQ2016a}]. Similar as
for the oxocuprate superconductors, there is a subtle interplay between
magnetism and superconductivity, and spin-fluctuations are considered as the
mechanism enabling the formation of Cooper pairs
[\onlinecite{Fernandes2013a}]. The possible coexistence of specific types
(spin-density wave, SDW) of magnetic order and superconductivity on the
microscale in the Fe-based systems is an ongoing topic of research.

The antiferromagnetic (afm) parent compounds of cuprate superconductors are
Mott (charge transfer) insulators. The situation is less clear for their
Fe-based counterparts. For instance, non-superconducting pnictides like
LaOFeAs or BaFe$_2$As$_2$ are metallic, itinerant antiferromagnets with low
ordered moments at the iron sites ($< 1 \mu_\mathrm{B}$), whereas
chalcogenides like K$_{0.8}$Fe$_{1.6}$Se$_2$ are insulators with large Fe
moments of $\approx 3 \mu_\mathrm{B}$. Strictly, however, half-filled systems
with $d^5$ configuration have to be regarded as the Mott-type parent
compounds, which is confirmed by the semiconducting properties of e.g.\
BaMn$_2$As$_2$
[\onlinecite{AnJ2009a,SinghY2009a,SinghY2009b,JohnstonDC2011a,ZhangWL2016a}]
as well as of LaOMnAs and LiMnAs [\onlinecite{Beleanu2013a,DongS2014a}].

There is an ongoing debate about the proper description of the electronic
structure of Fe-based pnictides and chalcogenides, in particular about the
importance of electron correlation and the degree of electron itinerancy
[\onlinecite{DAI12,MAN14}]. Since the basic Fe-As layers with formally
Fe$^{2+}$ ions in the parent compounds constitute a multi-orbital system,
orbital ordering is also believed to be of importance. Two opposing views have
been adopted: On the one hand, the iron pnictides have been considered as
weakly correlated metals where a SDW magnetic state is formed due to Fermi
surface nesting. On the other hand, in particular the more strongly correlated
selenides have been discussed as Mott-type insulators and the magnetism was
modeled in terms of Heisenberg-type exchange interactions between localized
moments. However, none of the limiting views can describe the many facets of
magnetism which have emerged from experimental studies and actually it is
believed that the Fe-based pnictides and chalcogenides are in between these
extremes. For instance, even itinerant electronic systems may show pronounced
correlation effects arising from Hund's rule coupling (Hund's metals
[\onlinecite{SiQ2016a,Val15}]), which rationalizes the existence of large
magnetic moments on short timescales of the order of femtoseconds, as derived
from photoemission experiments [\onlinecite{MAN14}].

An important prototype system is BaFe$_2$As$_2$ which adopts the tetragonal
ThCr$_2$Si$_2$-type crystal structure at room temperature and features a
concomitant structural and magnetic transition at $T_\mathrm{N} = 132$\,K
[\onlinecite{ROT08}]. Below $T_\mathrm{N}$ a stripe-like afm order occurs with
saturated Fe moments of about $0.9 \mu_\mathrm{B}$ being aligned along the $a$
axis of the low-temperature orthorhombic crystal structure
[\onlinecite{HUA08}]. BaFe$_2$As$_2$ can be converted into a
high-$T_\mathrm{c}$ superconductor by hole or electron doping
[\onlinecite{ROT08,CHE08,SAS08}] or by high pressure
[\onlinecite{Alireza2008a}]. While substitution of Fe by some transition
metals like Co, Ni or even Ru [\onlinecite{SEF08,LEI08,SCH09}] induces
superconductivity, substitution by others like Mn does not lead to
superconducting states [\onlinecite{KAS11}]. In contrast to BaFe$_2$As$_2$,
the half-filled Mn-analogue BaMn$_2$As$_2$ remains tetragonal down to low
temperatures and adopts a $G$-type afm spin structure with a high ordered
moment ($3.9 \mu_\mathrm{B}$) and a high ordering temperature of 625\,K
[\onlinecite{SinghY2009b,JohnstonDC2011a}]. Similarly high N\'eel temperatures
(692--758\,K) and $G$-type afm have been observed for the Cr species in the
isostructural silicides $R$Cr$_2$Si$_2$ ($R$ = Tb, Ho, Er)
[\onlinecite{MOZ03}].

Superconductivity in manganese and chromium compounds is very rare and only
recently representatives have been found. The first, binary CrAs, has been
long known for its structural and helimagnetic ordering transition at
240--190\,K with sizable Cr moments of $1.67 \mu_\mathrm{B}$
[\onlinecite{BOL71}]. The ordering temperature decreases dramatically to zero
for pressures of 0.7--0.8 GPa and superconductivity with critical temperature
up to 2.2\,K appears [\onlinecite{WuW2014a,Kotegawa2014a}]. At ambient
pressure superconductivity has been investigated in the new $A_2$Cr$_3$As$_2$
($A$ = K, Rb, Cs) compounds featuring quasi onedimensional Cr$_3$As$_3$ tubes
[\onlinecite{BaoJK2015a,CaoGH2016a}].

The Mott scenario for the transition-metal arsenides
[\onlinecite{IshidaLiebsch2010a}] considers a mirror-symmetry in the many-body
physics arising when the half-filled $d^5$ configuration is either doped by
electrons or by holes. The Cr compounds formally have a $d^4$ configuration
and in this sense are the hole-doped analogues to the electron-doped $d^6$
system BaFe$_2$As$_2$. Recently, some theoretical studies explored the
possibilities to find superconductivity for Ba$M_2$As$_2$ and LaO$M$As with
transition-metals $M$ (or mixtures) with less than 5 electrons
[\onlinecite{EDE16,PIZ16}]. Emphasis in these works is put on the strength of
correlations in dependence of the band filling. It is hoped that correlations
similar to those in the actual iron-based high $T_\mathrm{c}$ materials lead
to superconductivity.

Here, we focus on the system Ba(Fe$_{1-x}$Cr$_x$)$_2$As$_2$, where no
superconducting compositions have been found so far
[\onlinecite{KAS11,SEF09,Marty2011a}]. Theoretical calculations predicted an
afm checkerboard ($G$-type) ordering in the end member BaCr$_2$As$_2$
[\onlinecite{SinghDJ2009a}] and a ferromagnetic (fm) ground state in
atomically ordered BaCrFeAs$_2$ [\onlinecite{SinghY2009b}]. Another electronic
structure calculation predicted ordered BaCrFeAs$_2$ as a fully compensated
antiferromagnet with an iron moment of 2.6\,$\mu_\mathrm{B}$
[\onlinecite{HU10}]. Such a material could be relevant for spintronics
applications. Neutron diffraction studies on iron-rich
BaFe$_{2-x}$Cr$_x$As$_2$ single crystals with $0 < x < 0.94$ indicated that
near $x = 0.6$ the SDW ground state is replaced by a $G$-type afm state
[\onlinecite{Marty2011a}], but spin structure and ordering temperature of the
end member BaCr$_2$As$_2$ have not been studied experimentally yet. Also
electronic structure calculations of EuCr$_2$As$_2$ suggested a stable
$G$-type afm order of the Cr sublattice (in addition, the Eu$^{2+}$ ions show
fm ordering below 21\,K) [\onlinecite{PAR14}]. A recent experimental study
confirmed these predictions [\onlinecite{NAN16}].

We investigated the detailed crystal and magnetic structures of BaCr$_2$As$_2$
and BaCrFeAs$_2$ by temperature dependent powder neutron diffraction. Magnetic
susceptibility, electrical transport, specific heat, and $^{57}$Fe M\"ossbauer
spectroscopy measurements complement the study. Our theoretical studies
predict a random occupation of Cr and Fe on the transition-metal site for
BaCrFeAs$_2$ and metallic afm $G$-type ordered ground states for both
compounds. Experimentally, it is shown that both compounds in fact feature
$G$-type afm order, where the N\'eel temperature $T_\mathrm{N} = 580$\,K of
BaCr$_2$As$_2$ is nearly as high as that of BaMn$_2$As$_2$, although the
magnetic moment is only half as large. Both compounds are metallic conductors.
BaCrFeAs$_2$ turns out to be atomically disordered, and thus only average
magnetic moments are obtained from neutron diffraction. Most interestingly,
the Fe M\"ossbauer spectra indicate that the Fe magnetic moments are much
smaller than those of Cr and of similar size as in BaFe$_2$As$_2$.
$T_\mathrm{N}$ decreases with increasing iron content, thus the $G$-type order
is unfavorable for the Fe moments. The itinerant character of the magnetism
persists in the whole stability range of the $G$-type order, whereas
BaMn$_2$As$_2$ with the same nominal $d$ electron count as BaCrFeAs$_2$ is a
semiconductor.

%%%%%%%%%%%%%%%%%%%%%%%%%%%%%%%%%%%%%%%%%%%%%%%%%%%%%%%%%%%%%%%%%%%%%%%%%%%%%%%

\section{Experimental \& Calculation Details}
\label{sec:experimental}

\begin{figure}[htb]
\includegraphics[width=\columnwidth]{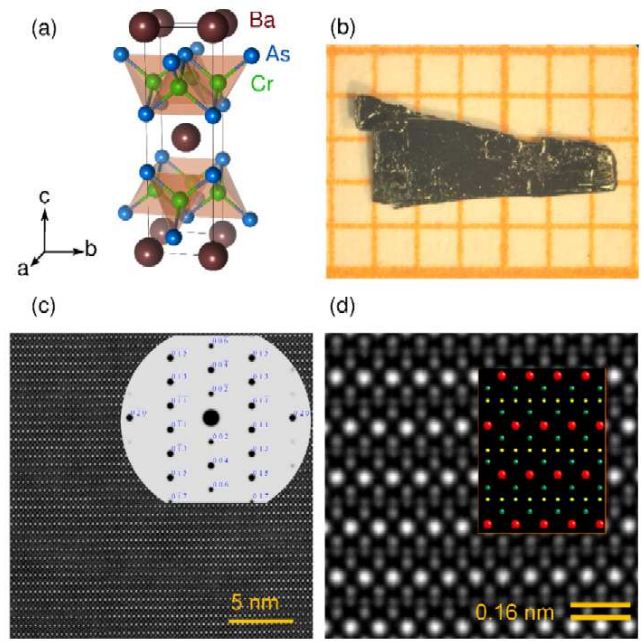}
\caption{a.) BaCr$_2$As$_2$ crystallizes in the ThCr$_2$Si$_2$ structure
(space group $I4/mmm$). b.) Crystals form as shiny black platelets. c.) HAADF
image and diffraction pattern. d.) Zoom of the HAADF image, the inset shows
the theoretical atom positions (Ba red, Cr green, As yellow).}
\label{fig:stru} \end{figure}

BaCr$_2$As$_2$ and BaCrFeAs$_2$ were synthesized by reacting CrAs and FeAs
with Ba in an alumina crucible. The binaries were synthesized according to
Singh \textit{et al}.\ [\onlinecite{SinghDJ2009a}] using Cr (chemPUR,
99.99\,\%), Fe (chemPUR, 99.9\,\%), and As (Alfa Aesar, 99.999\,\%). To
control the harsh reaction between the binaries and Ba (Alfa Aesar,
99.999\,\%), only 1/3 of the barium was added initially. The mixture was
heated to 1423\,K and held for 12\,h. Then the product was ground and
additional Ba was added. These steps were repeated until the powder x-ray
diffraction pattern (PXRD) did not show any CrAs or FeAs impurities. In both
cases an excess of $\approx 5$\,\% Ba was added to react all of the strongly
magnetic CrAs or FeAs. Single crystals were grown using the Bridgman technique
with alumina crucibles under argon atmosphere. The crucible was heated to
1823\,K and held for 24\,h, then it was moved with 1\,mm\,h$^{-1}$ out of the
hot zone of the furnace. Single crystals (Figure \ref{fig:stru}b) up to
10$\times$5$\times$4 mm could be grown. For the neutron diffraction
experiments $\approx 4$\,g of polycrystalline sample was prepared.

The crystal structure (ThCr$_2$Si$_2$ type, $tI10$, space group (SG) $I4/mmm$,
No.\ 139, Ba on $2a$, Cr on $4d$, As on $4e$) was confirmed by PXRD using
Cu-K$_{\alpha}$ radiation ($\lambda = 1.54056$\,\AA), with a Huber G670 camera
(Guinier technique). The refined lattice parameters are $a = 3.9667$ and $c =
13.6214$\,\AA\ for BaCr$_2$As$_2$ and $a = 3.986$ and $c = 13.2939$\,\AA\ for
BaCrFeAs$_2$. Additional reflexes suggest the presence of traces of
BaAl$_2$O$_4$ in some samples. High-resolution transmission electron
microscopy (HR-TEM) microstructure studies were performed using a FEI Titan
80-300. In the high-angle annular dark field (HAADF) images (shown in Figures
\ref{fig:stru}c and d for BaCr$_2$As$_2$; not shown for BaCrFeAs$_2$) no signs
of disorder on the Ba $2a$(0,0,0), Cr/(CrFe) $4d$
($\tfrac{1}{2},\tfrac{1}{2},\tfrac{1}{2}$) or As $4e$ (0,0,$z$) sites could be
found. In agreement with the PXRD pattern, segregates of BaAl$_2$O$_4$ were
seen on the surface for some BaCrFeAs$_2$ crystals. The average composition of
the crystals was found to be Ba$_{1.03(4)}$Cr$_{2.01(5)}$As$_{2.03(5)}$ and
Ba$_{1.01(6)}$Cr$_{0.96(8)}$Fe$_{1.08(2)}$As$_{2.10(1)}$ by chemical analysis
with inductively coupled plasma optical emission spectroscopy (ICP-OES).

Powder neutron diffraction experiments on BaCr$_2$As$_2$ and BaCrFeAs$_2$ have
been carried out on the instruments E2, E6, and E9 at the BER II reactor of
the Helmholtz-Zentrum Berlin. The instrument E9 uses a Ge-monochromator
selecting the neutron wavelength $\lambda = 1.7973$\,\AA, while the
instruments E2 and E6 use a pyrolytic graphite monochromator selecting the
neutron wavelength $\lambda = 2.38$\,{\AA} and 2.42\,\AA, respectively. On
these instruments powder patterns were recorded in the following ranges of
diffraction angles: $7.7^\circ \leq 2\theta \leq 83.4^\circ$ (E2), $7.5^\circ
\leq 2\theta \leq 136.5^\circ$ (E6), and $8^\circ \leq 2\theta \leq 141^\circ$
(E9). In case of BaCr$_2$As$_2$ powder patterns were collected between 2 and
750\,K, for BaCrFeAs$_2$ between 2 and 275\,K. The refinements of the crystal
and magnetic structure were carried out with FullProf [\onlinecite{ROD93}]
using the nuclear scattering lengths $b$(O) = 5.805\,fm, $b$(Cr) = 3.635\,fm,
$b$(Fe) = 9.54\,fm, $b$(Ba) = 5.25\,fm [\onlinecite{SEA95}]. The magnetic form
factors of the Cr$^{3+}$ and Fe$^{3+}$ ions were taken from Ref.\
\onlinecite{BRO95}.

Magnetization was measured in two magnetometer systems (MPMS-XL7 and MPMS3,
Quantum Design). Only small differences were seen between data taken in
warming after zero-field cooling (zfc) and during cooling in field (fc). For
some crystals such differences seem to be due to tiny CrAs precipitations (see
below). For one BaCr$_2$As$_2$ crystal the oven option of the MPMS3 was
employed. For these measurements the crystal was first measured at low
temperatures and then cemented to a heater stick for the high-temperature run.
The electrical resistivity, transverse magnetoresistance, Hall effect, and
heat capacity were determined with a measurement system (ACT and HC options,
respectively, PMMS9, Quantum Design) between 1.8 and 320\,K.

Temperature-dependent M\"ossbauer spectra of BaCrFeAs$_2$ were measured with a
standard WissEl spectrometer and a Janis closed cycle refrigerator (SHI-850-5)
using a $^{57}$Co/Rh source and a drive system operating in constant
acceleration mode. The M\"ossbauer absorber consisted of an acrylic glass
sample container with a sample density of $\approx 17$\,mg per cm$^2$. In
order to ensure homogeneous distribution the sample was mixed with boron
nitride. Spectra were collected between 5 and 292\,K and evaluated with
MossWinn [\onlinecite{MossWinn}]. Hyperfine field distributions were extracted
using the Hesse-R\"ubartsch method. Isomer shifts are given relative to
$\alpha$-iron.

%%%%%%%%%%%%%%%%%%%%%%%%%%%%%%%%%%%%%%%%%%%%%%%%%%%%%%%%%%%%%%%%%%%%%%%%%%%%%%%

The electronic structures were calculated self-consistently in the local spin
density approximation (LSDA) using the full potential linearized augmented
plane wave (FLAPW) method including spin-orbit (SO) interaction as implemented
in {\sc Wien}2k [\onlinecite{BSS99,SBl02,BSM01}]. For atomically disordered
BaCrFeAs$_2$ the calculations were carried out using the full potential, spin
polarized relativistic Korringa--Kohn--Rostocker method (SPRKKR)
[\onlinecite{Ebe99,EKM11}]. All calculations were performed using the PBE
exchange-correlation functional [\onlinecite{PBE96}] and the generalized
gradient approximation (GGA). To estimate the influence of a static
correlation, the LDA$+U$ method was applied.

For BaCr$_2$As$_2$ the calculations were started with the values $a = 3.96$,
$c = 13.6$\,\AA, and $z = 0.361$ [\onlinecite{PNa80}]. A $G$-type afm order
was assumed and the lattice was set up in SG $I\bar{4}m2$ (119) and the Cr
atoms were located on the split $2c$ and $2d$ sites to allow for the
anti-parallel orientations of their magnetization. This space group concerns
the atomic positions, the final magnetic symmetry depends on the direction of
the quantization axis and is described by a Shubnikov color group (here
$P\:4'/m'm'm$ (123.344)). A full structural optimization showed that the above
given parameters are nearly relaxed, in agreement with previous observations
[\onlinecite{SinghDJ2009a}].

In the calculations of the electronic structure for BaCrFeAs$_2$ the same
basic atomic and magnetic structures were used. Two cases were assumed, an
ordered and an alloyed structure. In the ordered structure, the initial $4d$
position of the Cr atom was split and half of the new positions was occupied
by Fe. This results in SG $I\bar{4}2m$ (119) with Ba $2a$, Cr $2c$, Fe $2d$,
and As $4e$. Starting from the experimental lattice parameters, an
optimization of the ordered structure resulted in $a_\mathrm{opt} =
4.012$\,\AA, $c_\mathrm{opt} = 13.3819$\,\AA, and $z_\mathrm{opt} = 0.357$.
The strong deviations of these optimized parameters from the experimental ones
suggest that the ordered case is not realized, i.\ e.\ in the actual material
Cr and Fe are disordered. For the alloyed structure it was assumed that Fe and
Cr occupy both the $2c$ and $2d$ position in a 50:50 ratio.

%%%%%%%%%%%%%%%%%%%%%%%%%%%%%%%%%%%%%%%%%%%%%%%%%%%%%%%%%%%%%%%%%%%%%%%%%%%%%%%

\section{Results}
\label{sec:results}

\subsection{Crystal structures}
\label{sec:stru}

\begin{table}
\caption{Results of the crystal structure refinements for BaCr$_2$As$_2$ and
BaCrFeAs$_2$. Powder neutron diffraction data were collected at $T = 2$\,K on
the instrument E9. The refinements of the data sets were carried out in the
tetragonal space group $I4/mmm$. The isotropic temperature factors of the
different atoms were constrained to be equal.}
\begin{ruledtabular}
\begin{tabular}{cccccc}
\multicolumn{6}{c}{BaCr$_2$As$_2$ at 2\,K, $R_\mathrm{N}$ = 0.0480} \\
\hline
\multicolumn{6}{c}{$a = 3.9503(2)$\,\AA, $c = 13.6047(10)$\,\AA, $Z = 2$, $V = 212.30(3)$\,\AA$^3$} \\
\hline
atom & site & $x$ & $y$            & $z$            & $B$ (\AA$^2$) \\
\hline
Ba   & $2a$ & 0   & 0              & 0              & 0.32(3)       \\
Cr   & $4d$ & 0   & $\tfrac{1}{2}$ & $\tfrac{1}{4}$ & 0.32(3)       \\
As   & $4e$ & 0   & 0              & 0.36092(14)    & 0.32(3)       \\[0.5ex]
\multicolumn{6}{c}{BaCrFeAs$_2$ at 2\,K, $R_\mathrm{N}$ = 0.0402} \\
\hline
\multicolumn{6}{c}{$a = 3.9793(1)$\,\AA, $c = 13.2532(4) $\,\AA, $Z = 2$, $V = 209.86(1)$\,\AA$^3$} \\
\hline
atom & site & $x$ & $y$            & $z$            & $B$ (\AA$^2$) \\
\hline
Ba   & $2a$ & 0   & 0              & 0              & 0.19(2)       \\
Cr   & $4d$ & 0   & $\tfrac{1}{2}$ & $\tfrac{1}{4}$ & 0.19(2)       \\
Fe   & $4d$ & 0   & $\tfrac{1}{2}$ & $\tfrac{1}{4}$ & 0.19(2)       \\
As   & $4e$ & 0   & 0              & 0.35779(10)    & 0.19(2)       \\
\end{tabular}
\end{ruledtabular}
\label{tab:stru}
\end{table}

\begin{figure}[htb]
\includegraphics[width=0.90\columnwidth]{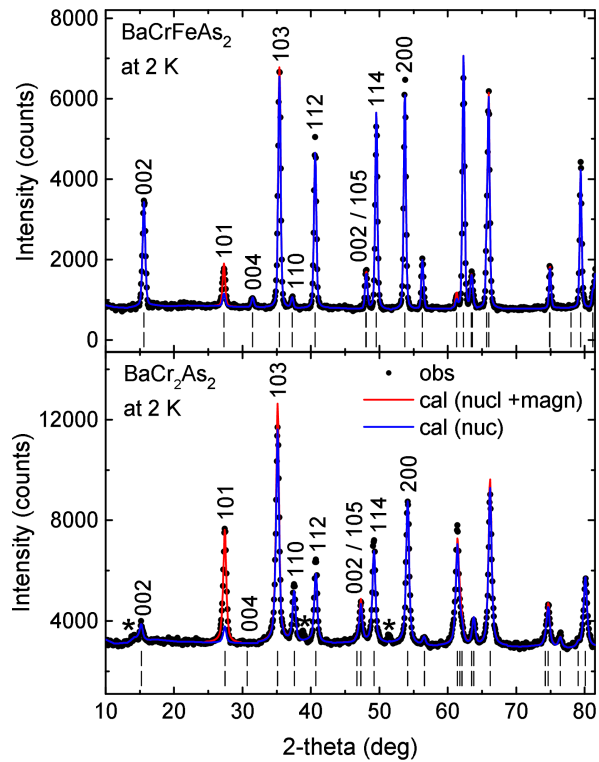}
\caption{Neutron diffraction patterns of BaCr$_2$As$_2$ and BaCrFeAs$_2$
powder taken at $T = 2$\,K. The calculated patterns of the pure nuclear
contribution (blue) as well as the sum of the nuclear and the magnetic
contribution (red) are compared with the observed ones (black full circles).
In the lower part the positions of the Bragg peaks are given (black ticks).
The BaCr$_2$As$_2$ sample contained an impurity (positions marked with
$\ast$), which could not be identified so far.} \label{fig:pattern}
\end{figure}

The crystal structures of BaCr$_2$As$_2$ and BaCrFeAs$_2$ were investigated by
powder neutron diffraction. Note that below $T_\mathrm{N} = 580$\,K,
respectively 265\,K, the patterns also included magnetic Bragg intensity (see
Section \ref{sec:magstru}). Both compounds crystallize in the tetragonal SG
$I4/mmm$ (No.\ 139), where the Ba, Cr(Fe), and As atoms occupy the Wyckoff
positions $2a$(0,0,0), $4d$(0,$\tfrac{1}{2}$,$\tfrac{1}{4}$), $4e$(0,0,z),
respectively. The same SG was reported earlier for BaMn$_2$As$_2$
[\onlinecite{SinghY2009a}]. From the data sets collected on the
fine-resolution powder diffractometer E9 we could not find any additional peak
splitting indicating a lower crystal structure symmetry. The refinements of
the crystal structure of BaCr$_2$As$_2$ from data sets recorded in the
temperature range from 2 up to 750\,K resulted in residuals between
$R_\mathrm{N} = 0.048$ and 0.088 [defined as $R_\mathrm{N} = (\Sigma
||F_\mathrm{obs}| - |F_\mathrm{calc}||) / \Sigma| F_\mathrm{obs}|$]. These
values are somewhat larger than expected. This can be ascribed to the fact,
that the investigated sample contained an impurity, which could not be
characterized so far (see Figure \ref{fig:pattern}). Further, additional
impurity peaks of the sample container were observed in the diffraction
patterns using the high temperature furnace on E9. Nevertheless, the lattice
parameters, as well as the positional parameters of BaCr$_2$As$_2$ could be
refined with good accuracy. In the case of BaCrFeAs$_2$ the crystal structure
was investigated in the temperature range between 2 and 275\,K. Due to the
higher purity of this sample the refinements resulted in smaller residuals
between $R_\mathrm{N} = 0.035$ and 0.043. The results of the Rietveld
refinements of the data sets collected at 2\,K are given in Table
\ref{tab:stru}. It is emphasized that there are no indications of
superstructure reflections which would point to atomic order of Cr and Fe
atoms.

\begin{figure}[htb]
\includegraphics[width=\columnwidth]{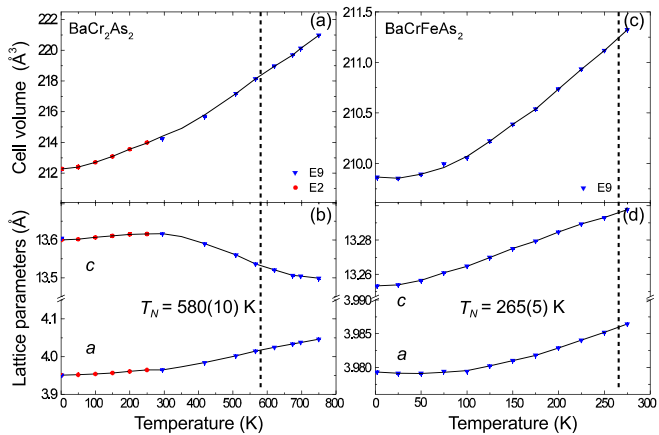}
\caption{Temperature dependence of the lattice parameters and the cell volumes
of BaCr$_2$As$_2$ and BaCrFeAs$_2$. These values were determined from data
sets collected on the instruments E2 and E9.} \label{fig:BCA_lattice}
\end{figure}

\begin{figure}[htb]
\includegraphics[width=\columnwidth]{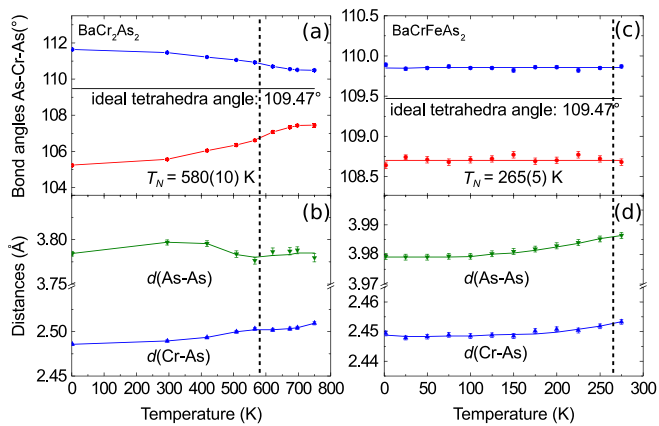}
\caption{Temperature dependence of the interatomic distances $d$(Cr-As)
[$d$(Cr/Fe-As)] and $d$(As-As) in BaCr$_2$As$_2$ and BaCrFeAs$_2$. Also shown
are the two bond angles $\angle_1$(As-Cr-As) (blue) and $\angle_2$(As-Cr-As)
(red) in the CrAs$_4$ tetrahedron of BaCr$_2$As$_2$ and the corresponding
angles for BaCrFeAs$_2$.} \label{fig:BCFA_angle} \end{figure}

Figure\ \ref{fig:BCA_lattice}a shows the temperature dependence of the cell
parameters as well as the cell volume. For BaCrFeAs$_2$ the lattice parameters
$a$ and $c$ show a continuous increase from 2 up to 275\,K, which is above the
magnetic ordering temperature $T_\mathrm{N} = 265(5)$\,K. A similar trend was
found for BaMn$_2$As$_2$ [\onlinecite{SinghY2009a}]. A continuous increase of
$a$ and $b$ was also observed for BaCr$_2$As$_2$ but only in the range from
2\,K up to about 300\,K. Further heating up to 700\,K leads to a decrease of
the $c$ parameter, while the increase of the $a$-parameter becomes somewhat
more pronounced (see Figure\ \ref{fig:BCA_lattice}). As a consequence, the
ratio $c/a$ decreases from 3.43 at 300\,K to 3.34 at 750\,K. The former value
is in agreement with literature data [\onlinecite{PNa80}]. These results
indicate that the magnetic transition in BaCr$_2$As$_2$ is accompanied by
subtle structural modifications, in contrast to BaMn$_2$As$_2$
[\onlinecite{SinghY2009a}]. The anomaly in the $c$ parameter is reflected in
the change of the distance between the arsenic atoms lying along the $c$
direction. Namely, one observes a decrease of the distance $d$(As-As) between
about 300\,K and the N\'eel temperature $T_\mathrm{N} = 580(10)$\,K as shown
in Fig.\ \ref{fig:BCFA_angle}. Above $T_\mathrm{N}$ finally $d$(As-As) seems
to show a slight increase up to 750\,K. The bond length $d$(Cr-As) only shows
a slight increase from 2\,K up to 750\,K, but a weak anomaly is observed near
$T_\mathrm{N}$. On the other hand in BaCrFeAs$_2$ both, $d$(As-As) and
$d$(Cr-As), only show a slight increase in the temperature range from 2\,K and
275\,K without anomalies, indicating that magnetic ordering does not influence
the structural properties significantly. In Fig.\ \ref{fig:BCFA_angle} we have
also plotted the changes of the two different bond angles $\angle_1$(As-Cr-As)
and $\angle_2$(As-Cr-As) in the CrAs$_4$ (and Cr/FeAs$_4$) tetrahedra, which
form a twodimensional layer in the $ab$ plane. The structural anomalies
accompanying the magnetic transition in BaCr$_2$As$_2$ are also reflected in
the bond angles. Thus, the bond angles $\angle_1$(As-Cr-As) and
$\angle_2$(As-Cr-As) show a strong increase and decrease in the magnetically
ordered state down to 2\,K. Above 650\,K both bond angles seem to reach
saturation values. In contrast, the two bond angles for BaCrFeAs$_2$ are
practically unchanged between 2 and 275\,K. However, both pnictides do not
reach the ideal tetrahedron angle of $109.47^\circ$. The CrAs$_4$ tetrahedra
in BaCr$_2$As$_2$ are stronger distorted than the Cr/FeAs$_4$ tetrahedra in
BaCrFeAs$_2$.

%%%%%%%%%%%%%%%%%%%%%%%%%%%%%%%%%%%%%%%%%%%%%%%%%%%%%%%%%%%%%%%%%%%%%%%%%%%%%%%

\subsection{Electronic structure of BaCr$_2$As$_2$}
\label{sec:theoryBCA}

\begin{figure}[htb]
\includegraphics[width=0.70\columnwidth]{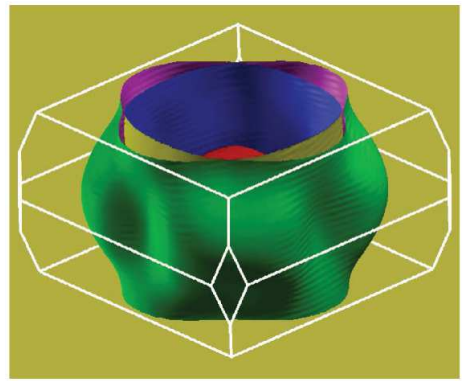}
\caption{Fermi surface of BaCr$_2$As$_2$.} \label{fig:LAPW-FS} \end{figure}

In Figure \ref{fig:LAPW-FS} the Fermi surface of BaCr$_2$As$_2$ is shown.
Three bands are crossing the Fermi energy $\epsilon_\mathrm{F}$, resulting in
three iso-surfaces. The innermost one is closed and has a cushion shape. It is
related to a hole pocket around the $\Gamma$ point. The outer two are open at
top and bottom and have corrugated cylinder shapes with fourfold arranged
bulges. The effective band mass $m^\star/m_e$ of the innermost hole pocket was
determined for the $\Delta$ and $\Sigma$ directions revealing values of
$m^\star_\Delta = -0.63$ and $m^\star_\Sigma = -0.39$ at the $\Gamma$ point.
This hints that this band is in the GGA-LSDA not responsible for the observed
enhanced value of the electronic specific heat coefficient $\gamma$ (cf.\
Ref.\ \onlinecite{SinghDJ2009a} and our specific heat results in section
\ref{sec:mag}).

\begin{figure}[htb]
\includegraphics[height=1.10\columnwidth]{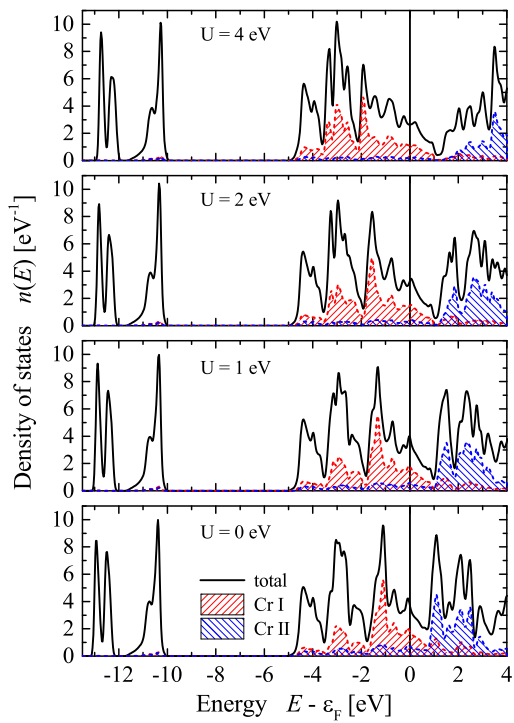}
\caption{Density of states of BaCr$_2$As$_2$ calculated with LDA$+U$. Shown is
the total DOS together with the local DOS at Cr atoms for increasing $U$.}
\label{fig:dos} \end{figure}

The evolution of the DOS with increasing Coulomb parameter $U$ in the LDA$+U$
calculation is given in Figure \ref{fig:dos}. Note that the unoccupied
minority states of one of the Cr atoms correspond to the unoccupied majority
states of the other, and vice versa for the occupied states due to the
anti-symmetric spin densities of an antiferromagnet. The increased $U$ results
in an increase of the magnetic moment at the Cr atoms (cf.\ Table
\ref{tab:BCA}). Their values are in the range 2.4-3.5\,$\mu_\mathrm{B}$.

The ``bare'' coefficient of the electron specific heat $\gamma_0 = \pi^2 k_B^2
n(\epsilon_\mathrm{F})/3$ for BaCr$_2$As$_2$ is $\approx
8$\,mJ\,mol$^{-1}$K$^{-2}$. For the various calculation schemes the results
are also given in Table \ref{tab:BCA}. The inclusion of SO interaction has
only a weak effect on $\gamma_0$. For LDA$+U$, it is found that
$n(\epsilon_\mathrm{F})$ -- and thus $\gamma_0$ -- increases slightly and has
a maximum at $U \approx 1$\,eV, but then decreases with further increase of
$U$. Using SPRKKR and the mean field approximation a N\'eel temperature
$T_\mathrm{N,calc} \approx 880$\,K can be calculated.

\begin{table}[htb]
\caption{Spin magnetic moment of the Cr atoms, the DOS at the Fermi energy
$n(\epsilon_\mathrm{F})$, and the coefficient of the electron specific heat
$\gamma_0$ for different calculation schemes for $G$-type afm ordered
BaCr$_2$As$_2$. All calculations were performed including spin-orbit
interaction (SO).}
\begin{ruledtabular}
\begin{tabular}{l cccc}
            & $U$ & $m_\mathrm{Cr}$  & $n(\epsilon_\mathrm{F})$ & $\gamma_0$               \\
method      & eV  & $\mu_\mathrm{B}$ & eV$^{-1}$                & mJ\,mol$^{-1}$\,K$^{-2}$ \\
\hline
GGA with SO & 0   & 2.4              & 3.37                     & 7.95                     \\
\hline
LDA$+U$     & 1   & 2.8              & 3.67                     & 8.65                     \\
            & 2   & 3.1              & 2.45                     & 5.77                     \\
            & 3   & 3.3              & 2.91                     & 6.85                     \\
            & 4   & 3.4              & 2.85                     & 6.71                     \\
            & 5   & 3.5              & 1.84                     & 4.35                     \\
\end{tabular}
\end{ruledtabular}
\label{tab:BCA}
\end{table}

%%%%%%%%%%%%%%%%%%%%%%%%%%%%%%%%%%%%%%%%%%%%%%%%%%%%%%%%%%%%%%%%%%%%%%%

\subsection{Electronic structure of BaCrFeAs$_2$}
\label{sec:theoryBCFA}

The electronic structure of the ordered variant of BaCrFeAs$_2$ should be a
completely compensated half-metallic ferromagnet as reported in Reference
[\onlinecite{HU10}]. The peculiarity of this type of magnetism is that the
moment of two different types of atoms -- here Cr and Fe -- compensate each
other similar to a antiferromagnet. The different symmetry, however, allows
that the DOS in the two spin channels are different. Moreover, a band gap at
$\epsilon_\mathrm{F}$ appears in one of the spin densities similar to a
half-metallic ferromagnet. The DOS for the $\uparrow$ spin channel has a
pronounced maximum just above $\epsilon_\mathrm{F}$ with a value of
$n(\epsilon_\mathrm{F}+\delta) = 6.92$\,eV$^{-1}$ ($\delta \approx 50$\,meV).
This value is twice as high as the one calculated for BaCr$_2$As$_2$. It
arises from flat bands around $\epsilon_\mathrm{F}$. A detailed analysis of
the DOS reveals that those states are localized at the Fe atoms. In many
cases, compounds with such a peaked DOS at $\epsilon_\mathrm{F}$ are not
stable.

\begin{figure}[htb]
\includegraphics[width=\columnwidth]{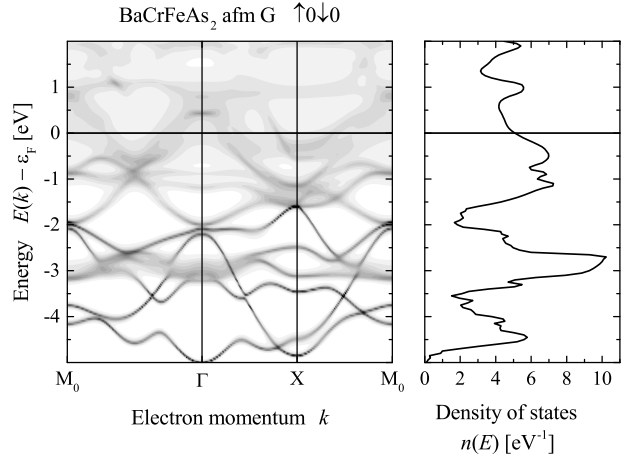}
\caption{Electronic structure of BaCrFeAs$_2$ with chemical disorder between
Cr and Fe. Shown is the case with a vanishing magnetic moment at Fe. The DOS
is the same in both spin channels due to the afm order, thus the sum is
plotted.} \label{fig:bcfa_alloy} \end{figure}

The electronic structure and the DOS of the alloy variant of BaCrFeAs$_2$ are
shown in Fig.\ \ref{fig:bcfa_alloy}. The result of the chemical disorder
scattering are strongly broadened electronic states, in particular close to
$\epsilon_\mathrm{F}$. The broadening causes a reduction of the maximum of the
DOS just above $\epsilon_\mathrm{F}$ to about 5\,eV$^{-1}$, which is still
higher compared to BaCr$_2$As$_2$.

The magnetic properties of ordered and alloyed BaCrFeAs$_2$ are compared in
Table \ref{tab:BCFA}. The moment for the atomically ordered phase compares
well with 2.6\,$\mu_\mathrm{B}$ reported in Ref.\ \onlinecite{HU10}. The
moments in the ordered case are not completely the same, the vanishing total
moment is guaranteed by the polarization of the interstitial and the atoms in
the vicinity of the magnetic Cr and Fe atoms. The orbital magnetic moments are
very small, in the order of 0.002\,$\mu_\mathrm{B}$ or 0.06\,$\mu_\mathrm{B}$.

Interestingly, for the random distribution of Fe and Cr, a stable solution is
found with vanishing moments at the Fe atoms at the same site. The calculation
was started with a starting moment at the Cr atoms ($4\mu_\mathrm{B}$) but
zero moment at Fe. No magnetic moment was induced during the self consistent
cycles. The result is an average moment of $1.38\mu_\mathrm{B}$ for the
$2c$/$2d$ positions.

\begin{table}[htb]
\caption{Magnetic properties of BaCrFeAs$_2$. Tabulated are the DOS just above
$\epsilon_\mathrm{F}$ ($n_\mathrm{max}$), the absolute values of the magnetic
moments at Cr and Fe species ($m_\mathrm{Cr/Fe}$), and the calculated N\'eel
temperatures ($T_\mathrm{N,calc}$).}
\begin{center}
\begin{ruledtabular}
\begin{tabular}{llll}
                    & ordered & disordered &                    \\
\hline
$n_\mathrm{max}$    & 6.92    & 4.99       & (eV$^{-1}$)        \\
$m_\mathrm{Cr} $    & 2.74    & 2.76       & ($\mu_\mathrm{B}$) \\
$m_\mathrm{Fe} $    & 2.53    & 0.00       & ($\mu_\mathrm{B}$) \\
$T_\mathrm{N,calc}$ & 965     & 260        & (K)                \\
\end{tabular}
\end{ruledtabular}
\end{center}
\label{tab:BCFA}
\end{table}

Also for BaCrFeAs$_2$ the N\'eel temperature can be calculated in the mean
field approximation using SPRKKR. The value for the atomically ordered
structure is very high, however, a much lower N\'eel temperature of about
260\,K is obtained for the disordered version.

%%%%%%%%%%%%%%%%%%%%%%%%%%%%%%%%%%%%%%%%%%%%%%%%%%%%%%%%%%%%%%%%%%%%%%%%%%%%%%%

\subsection{Magnetic and thermal properties, electrical transport}
\label{sec:mag}

\begin{figure}[htb]
\includegraphics[height=\columnwidth,angle=90]{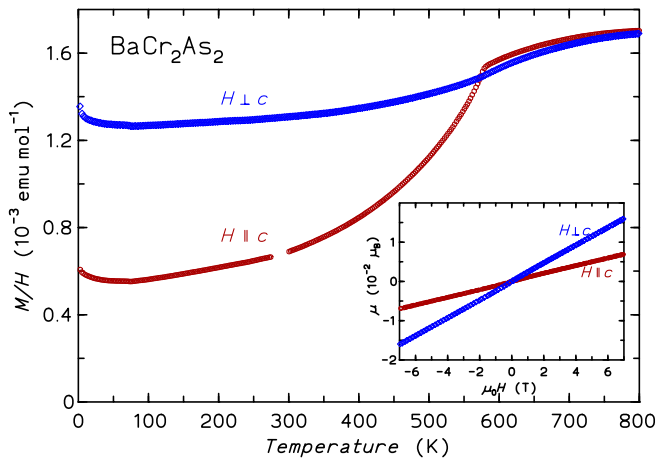}
\caption{Magnetic susceptibility of a BaCr$_2$As$_2$ crystal in a field
$\mu_0H = 1$\,T perpendicular (blue diamonds) and parallel (red circles) to
the crystallographic $c$ axis. Minor adjustments of the magnetometer oven data
sets ($T > 400\,K$) were made to match with the data below 400\,K. The inset
shows the magnetic moment from an isothermal magnetization loop measured at a
temperature $T = 1.8$\,K.} \label{fig:suscBCA} \end{figure}

The temperature dependence of the magnetic susceptibility, $\chi = M/H$, at
$\mu_0H = 1$\,T of a BaCr$_2$As$_2$ crystal platelet ($ab$ plane) was measured
with the magnetic field parallel and perpendicular to the $c$ axis (Fig.\
\ref{fig:suscBCA}). The small difference between the directions above 600\,K
suggests a rather weak anisotropy in the paramagnetic regime. However, no
Curie-Weiss behavior could be observed up to $T = 800$\,K. For $H \parallel c$
a clear kink is visible, indicating the afm ordering of the Cr sub-lattice at
$T_\mathrm{N} = 575(10)$\,K, as determined from the peak in the derivative
d($\chi T$)/d$T$. The decrease of $\chi(T)$ for $H \parallel c$ towards low
temperatures is much stronger than for $H \perp c$, eventually suggesting an
afm order with magnetic moments aligned along the $c$ axis. The behavior of
$\chi(T)$ and the ordering temperature is very similar to that of
BaMn$_2$As$_2$ [\onlinecite{JohnstonDC2011a}]. The nearly isotropic but
non-Curie-Weiss behavior above $T_\mathrm{N}$ is observed for several afm
compounds with ThCr$_2$Si$_2$-type structure. It is typical for the somewhat
twodimensional character of the magnetic interactions and has been treated in
detail theoretically for BaMn$_2$As$_2$ [\onlinecite{JohnstonDC2011a}]. For
BaCr$_2$As$_2$ the smooth maximum of $\chi(T)$ may be anticipated for
temperatures around 900\,K.

A small upturn of $\chi(T)$ in both directions for temperatures below $\approx
50$\,K suggests the presence of paramagnetic impurities, e.g.\ from point
defects. Magnetization loops taken at $T = 1.8$\,K (Fig.\ \ref{fig:suscBCA}
inset) show a very small fm-like component of $< 5 \times 10^{-4}
\mu_\mathrm{B}$ for both field along or perpendicular to the $c$ axis. Also,
in measurements of $M(T)$ at $\mu_0H = 0.01$\,T, a small sharp transition from
a fm impurity phase becomes visible at $T_{imp} \approx 76$\,K. We assign this
transition and the fm-like signals to the structural and helimagnetic ordering
transition of CrAs [\onlinecite{BOL71}]. The ordering temperature of CrAs is
known to decrease dramatically to zero for pressures of 0.7-0.8\,GPa
[\onlinecite{WuW2014a,Kotegawa2014a}]. We speculate that CrAs on the crystal
surface is strained and its ordering temperature is therefore reduced to
$T_\mathrm{imp}$.

\begin{figure}[htb]
\includegraphics[height=\columnwidth,angle=90]{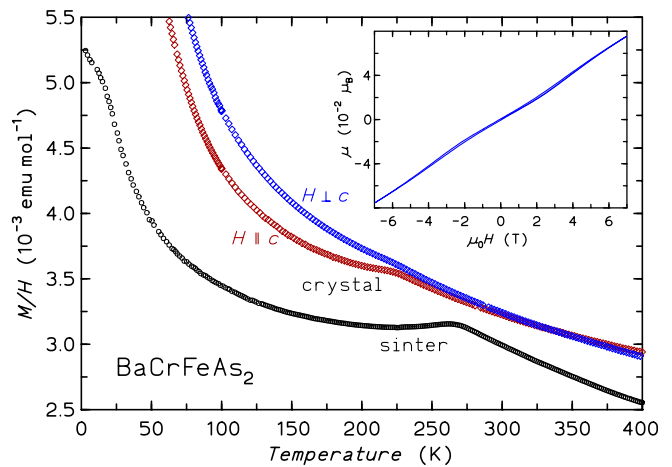}
\caption{Magnetic susceptibility of the BaCrFeAs$_2$ sintered sample used for
the other measurements (black circles) in a field $\mu_0H = 1$\,T. For
comparison, the susceptibility of a BaCrFeAs$_2$ crystal (diamonds) with field
perpendicular and parallel to the $c$ axis is shown (shifted up by $5 \times
10^{-4}$ emu mol$^{-1}$). The crystal has a lower N\'eel temperature. The
inset shows the magnetization loop at $T = 2.0$\,K} \label{fig:suscBCFA}
\end{figure}

The magnetic susceptibility of sintered BaCrFeAs$_2$ (Fig.\
\ref{fig:suscBCFA}) displays a broadened cusp at $T_\mathrm{N} \approx
269(2)$\,K (midpoint of step in d($\chi T$)/d$T$), but there is only a slight
decrease of the $\chi(T)$ curve below this temperature. Instead, with
decreasing temperature the susceptibility increases again, indicating a strong
paramagnetic contribution following a Curie law. Above the N\'eel temperature
(range 300-400\,K) the data are well fitted by a Curie-Weiss law with
effective moment $\mu_\mathrm{eff} = 3.70\,\mu_\mathrm{B}$ and Weiss
temperature $\theta_\mathrm{P} = -273$\,K. The BaCrFeAs$_2$ crystal shows a
similar cusp, however at $\approx 225$\,K. The lower $T_\mathrm{N}$ of the
single crystal is probably due to a slightly lower Fe content compared to the
sinter sample. The decrease of $\chi(T)$ is more pronounced for $H \parallel
c$, suggesting an afm ordered structure with the magnetic moments lying in the
crystallographic $c$ direction. The paramagnetic contribution in the
magnetically ordered state of BaCrFeAs$_2$ is discussed in connection with the
M\"ossbauer results (Sec.\ \ref{sec:mos}).

\begin{figure}[htb]
\includegraphics[height=\columnwidth,angle=90]{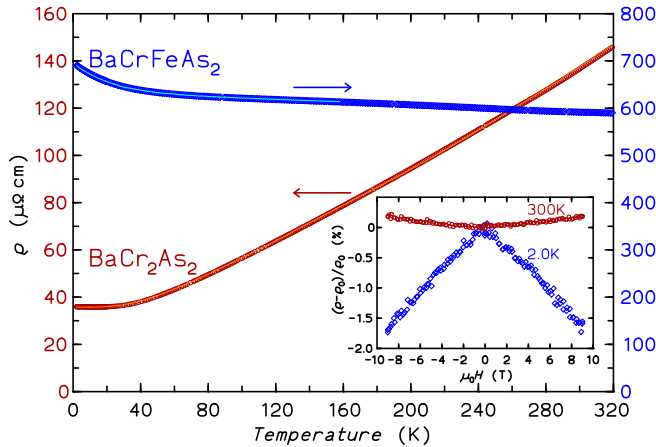}
\caption{Electrical resistivity of a BaCr$_2$As$_2$ crystal measured in the
crystallographic $ab$ plane (red circles) and of a sintered sample of
BaCrFeAs$_2$ (blue diamonds) in zero magnetic field. The resistivity curves in
$\mu_0H = 9$\,T (lines) almost coincide with the zero-field data. The inset
shows the magnetoresistance $100 \times (\rho - \rho_0)/\rho_0$ as function of
field for BaCr$_2$As$_2$ measured at $T = 2.0$ and 300\,K.} \label{fig:rho}
\end{figure}

\begin{figure}[htb]
\includegraphics[height=\columnwidth,angle=90]{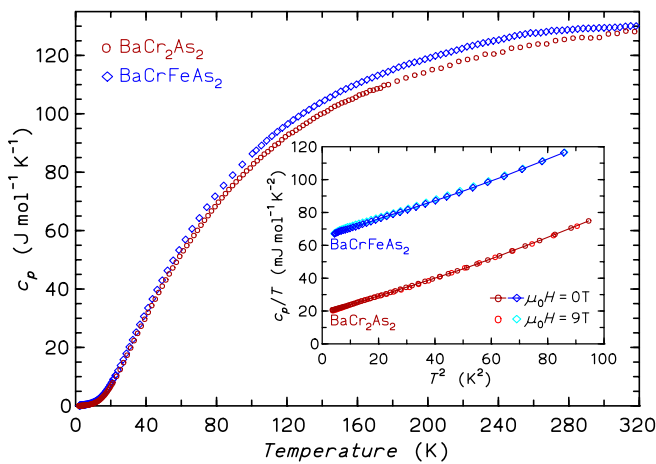}
\caption{Molar specific heat $c_p(T)$ of a BaCr$_2$As$_2$ single crystal (red
circles) and of polycrystalline sintered BaCrFeAs$_2$ (blue diamonds). Inset:
$c_p/T$ vs.\ $T^2$ representation for low temperatures. Zero-field data (dark
symbols and line) and data in a magnetic field $\mu_0H = 9$\,T (light symbols)
are shown.} \label{fig:cp} \end{figure}

The electrical resistivity, $\rho(T)$, of the BaCr$_2$As$_2$ crystal measured
for current in the $ab$ plane (Fig.\ \ref{fig:rho}) increases almost linearly
with temperature and indicates metallic behavior of the compound
($\rho_\mathrm{300\,K} = 136$\,$\mu\Omega$\,cm). The residual resistance ratio
RRR $\approx 3.8$ indicates a fair quality of the specimen and the residual
resistance is already reached at $\approx 13$\,K. Interestingly, below this
temperature $\rho(T)$ increase very slightly (by $\approx 0.1$\,\% of
$\rho_0$). The magnetoresistance (MR, Fig.\ \ref{fig:rho} inset) is very small
and positive for high temperatures. Surprisingly, for $T = 2.0$\,K it is
negative and quite strong (-1.7\,\% at $\mu_0H = 9$\,T). This might indicate
the damping of an additional scattering mechanism at very low temperatures
compared to the $T_\mathrm{N}$. The origin of both the upturn of $\rho(T)$
below 13\,K and of the negative MR might be due to a Kondo-hole effect.
Non-magnetic atoms replacing magnetic Cr in the afm ordered lattice can give
rise to a Kondo-like upturn of resistivity at low temperatures. This
scattering is weakened by the application of a magnetic field resulting in a
negative MR proportional to the field. The Hall resistivity curves
$\rho_{ab}(H)$ are linear, the Hall constants are positive and do almost not
vary with temperature. This observation is consistent with the hole pocket
around the $\Gamma$ point (cf.\ Fig.\ \ref{fig:LAPW-FS}). Within a one-band
model the Hall constant corresponds to a hole density $n_\mathrm{h} \approx
1.5 \times 10^{22}$\,cm$^{-3}$ and low mobility ($b_\mathrm{h} =
11$\,cm$^2$\,V$^{-1}$\,s$^{-1}$ at $T = 2$\,K).

For polycrystalline BaCrFeAs$_2$ (Fig.\ \ref{fig:rho}) as well as for a single
crystal (not shown) the electrical resistivity is about five time higher.
Interestingly, it displays only a weak temperature dependence and increases
continuously with decreasing temperature. Similar to BaCr$_2$As$_2$ there is
an upturn below a certain temperature (here, $\approx 40$\,K). The absence of
a typical metallic resistivity behavior is probably due to scattering of
charge carriers on the disordered Cr/Fe species of the $4d$ Wyckoff site. The
order of magnitude suggests however that BaCrFeAs$_2$ is still a metal.
Nominally isoelectronic BaMn$_2$As$_2$ crystals grown in MnAs flux show an
about 100 times higher resistivity [\onlinecite{SinghY2009a}]. The
magnetoresistance of BaCrFeAs$_2$ is small ($< 0.15$\,\%) for all
temperatures, indicating that magnetic scattering due to the atomic Cr/Fe
disorder is unimportant. Also for BaCrFeAs$_2$ the Hall isotherms
$\rho_{xy}(H)$ are linear, however Hall constants are negative and vary with
temperature. They correspond to electron densities $n_e = 8.3 \times
10^{21}$\,cm$^{-3}$ at $T = 300$\,K and only $1.5 \times 10^{21}$\,cm$^{-3}$
at 2\,K with $b_e = 6$\,cm$^2$\,V$^{-1}$\,s$^{-1}$.

The specific heat, $c_p(T)$, of the two compounds is presented in Fig.\
\ref{fig:cp}. For BaCr$_2$As$_2$ no anomalies from transitions are visible up
to 320\,K. However, $c_p(T)$ for BaCrFeAs$_2$ is larger than that of
BaCr$_2$As$_2$ in the covered temperature range which might also be due to the
Cr/Fe disorder. A very small steplike anomaly can be seen at $\approx 260$\,K
(barely standing out of the noise). The transition temperature is in agreement
with our magnetization and neutron diffraction data (see Section
\ref{sec:magstru}). The small entropy change connected with this magnetic
ordering is due to the low ordered moments of Cr/Fe and the predominantly
itinerant character of the magnetic system. The Dulong-Petit limit $c_p = 3nR$
($R$ = molar gas constant, $n$ = number of atoms) is reached by both compounds
at around room temperature.

At low temperatures the specific heats (see Fig.\ \ref{fig:cp} inset) are well
described by $c_p(T) = \gamma T + \beta T^3 + \delta T^5$ where $\gamma$ is
the coefficient of the linear term $\gamma T$ (usually assigned to conduction
electrons) and $\beta T^3$ and $\delta T^5$ are the first two terms due to the
harmonic theory of lattice specific heat. For BaCr$_2$As$_2$ (BaCrFeAs$_2$)
least-squares fits in the range 1.9--7\,K result in $\gamma = 18.8$ (64.9) mJ
mol$^{-1}$ K$^{-1}$, $\beta = 0.51$ (0.56) mJ mol$^{-1}$ K$^{-4}$
corresponding to an initial Debye temperature of 268 (259)\,K, and $\delta =
0.6$ (0.0) $\mu$J mol$^{-1}$ K$^{-6}$. The $\gamma$ value obtained for
BaCr$_2$As$_2$ is close to the value deduced by Singh \textit{et al}.\
[\onlinecite{SinghDJ2009a}], however the linear term of BaCrFeAs$_2$ is very
large. The $\gamma$ value for BaCrFeAs$_2$ is confirmed through measurements
on the single crystal for which we show the susceptibility in Fig.\
\ref{fig:suscBCFA}. The lattice properties (Debye temperatures) of the two
compounds are similar, as expected from the small atomic mass difference of Cr
and Fe. Also, afm spin waves do not play a role due to the high N\'eel
temperatures.

The linear specific heat contribution is insensitive to magnetic fields (cf.\
Fig.\ \ref{fig:cp} inset). For BaCr$_2$As$_2$ in a field $\mu_0H = 9$\,T the
linear coefficient $\gamma$ is not changed at all. For BaCrFeAs$_2$ the
specific heat increases slightly in $\mu_0H = 9$\,T for $T < 5$\,K (by
maximally 1.8\,\% at 3.3\,K), which might be due to a Schottky-type anomaly
due to minor impurities. Thus, the enhanced $\gamma$ values should either not
be due to spin fluctuations at all or the spin fluctuations are of too high in
energy. Especially BaCr$_2$As$_2$ has a very high N\'eel temperature and the
fluctuations might not be quenchable by such a small field.

%%%%%%%%%%%%%%%%%%%%%%%%%%%%%%%%%%%%%%%%%%%%%%%%%%%%%%%%%%%%%%%%%%%%%%%%%%%%%%%

\subsection{Magnetic structures}
\label{sec:magstru}

\begin{figure}[htb]
\includegraphics[width=0.9\columnwidth]{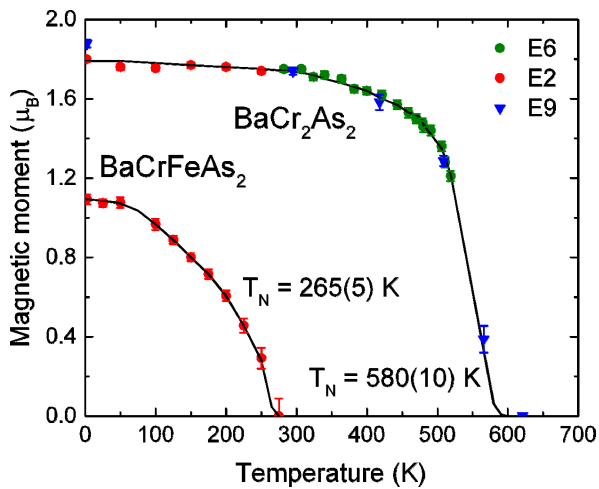}
\caption{Temperature dependence of the magnetic moments per transition-metal
atom in BaCr$_2$As$_2$ and BaCrFeAs$_2$ as obtained from powder neutron
diffraction data (instruments E2, E6, and E9). Magnetic intensity of Bragg
reflections disappears at the N\'eel temperatures $T_\mathrm{N} = 580(10)$\,K
and $T_\mathrm{N} = 265(5)$\,K respectively.} \label{fig:Crmoment}
\end{figure}

In order to investigate the magnetic structure of BaCr$_2$As$_2$ we have
collected a powder neutron pattern at 2\,K. In comparison to the data
collected at 750\,K, well above the N\'eel temperature $T_\mathrm{N} =
580$\,K, it was shown that magnetic intensities appear at the positions of
nuclear Bragg reflections indicating a magnetic structure, which can be
described with the propagation vector $\bm{k} = 0$. This shows that the
translation $\bm{t} = (\tfrac{1}{2},\tfrac{1}{2},\tfrac{1}{2})$ associated
with the $I$ cell is not lost. In Figure \ref{fig:pattern} it can be seen that
the strongest magnetic intensity is observed at $2\theta = 27.4^\circ$, which
is the position of the Bragg reflection 101. Strong intensity could be
generated at this position using a magnetic structure model, where the
chromium atoms in the Wyckoff positions $4d$, located at
$(0,\tfrac{1}{2},\tfrac{1}{4})$ and $(\tfrac{1}{2}$,0,$\tfrac{1}{4})$, are
coupled antiparallel. Due to the $I$ centering the operation $\bm{t} =
(\tfrac{1}{2},\tfrac{1}{2},\tfrac{1}{2})$ does not change the direction of the
spin. Therefore one finds the spin sequence $+-+-$ for the Cr atoms in the
positions $(0,\tfrac{1}{2},\tfrac{1}{4})$, $(\tfrac{1}{2},0,\tfrac{1}{4})$,
$(\tfrac{1}{2},0,\tfrac{3}{4})$, and $(0,\tfrac{1}{2},\tfrac{3}{4})$.

Using this structure model the magnetic structure could be successfully
refined, when the magnetic moments are aligned parallel to the $c$ axis. The
same type of magnetic ordering is found for BaMn$_2$As$_2$ and also for the Cr
sublattice in EuCr$_2$As$_2$ [\onlinecite{NAN16}]. Assuming an additional
magnetic component within the $ab$ plane, magnetic intensity should appear at
the position of the reflection 002. Magnetic intensity could not be easily
determined from the difference patterns collected at 2 and 750\,K, in the
fully ordered and the paramagnetic state, because the strong structural
changes lead to a change of the nuclear intensity of the reflection 002.
Therefore we have refined simultaneously the crystal and the magnetic
structure at 2\,K. The calculated nuclear intensity of the 002 reflection is
even slightly larger than the observed one. This clearly indicates the absence
of an additional magnetic component within the $ab$ plane, i.\ e.\ the absence
of any spin canting to the $c$ axis. From the data set collected at 2\,K on
instrument E9 we finally found for the Cr atoms a magnetic moment of
$\mu_\mathrm{exp} = 1.88(2) \mu_\mathrm{B}$ resulting in a residual
$R_\mathrm{M} = 0.0268$ [defined as $R_\mathrm{M} = (\Sigma ||I_\mathrm{obs}|
- |I_\mathrm{calc}||) / \Sigma |I_\mathrm{obs}|$]. From the data set collected
on instrument E2 the magnetic moment $\mu_\mathrm{exp} = 1.80(2)
\mu_\mathrm{B}$ was found to be slightly smaller (residual $R_\mathrm{M} =
0.118$). It has to be mentioned that the moment value determined from the E9
data is more reliable, since the overall scale factor could be determined with
higher accuracy (a much larger number of nuclear Bragg reflections was
available). From the data collected on E2, E6, and E9 we are able to determine
the temperature dependence of the magnetic moment up to the N\'eel temperature
$T_\mathrm{N} = 580(10)$\,K (see Figure \ref{fig:Crmoment}). The results from
the neutron diffraction experiments compare well with those from magnetization
data of the BaCr$_2$As$_2$ single crystal (Fig.\ \ref{fig:suscBCA}). The
experimental magnetic Cr moments at 2\,K compare reasonably well with the
values obtained from our (cf.\ Table \ref{tab:BCA} and previous
[\onlinecite{SinghDJ2009a}] electronic structure calculations. The
experimental Cr moment is even slightly smaller than the calculated moments.
This shows that electronic correlation effects reflected in the parameter $U$
of the LDA$+U$ calculations do not play a role as increasing $U$ lead to
larger Cr magnetic moments.

Accordingly we have determined the magnetic structure of BaCrFeAs$_2$. As it
can be seen in Figure \ref{fig:pattern} the strongest magnetic intensity could
also be observed at the position of the reflection 101. The magnetic
intensities of BaCrFeAs$_2$ are found to be much weaker than those in
BaCr$_2$As$_2$ indicating that the averaged magnetic moments of the Cr and Fe
atoms are strongly reduced. A moment $\mu_\mathrm{exp} = 1.09(3)
\mu_\mathrm{B}$ was refined resulting in a residual $R_\mathrm{M} = 0.068$.
The N\'eel temperature $T_\mathrm{N} = 265(5)$\,K compares well with our
magnetic susceptibility data (Fig.\ \ref{fig:suscBCFA}).

%%%%%%%%%%%%%%%%%%%%%%%%%%%%%%%%%%%%%%%%%%%%%%%%%%%%%%%%%%%%%%%%%%%%%%%%%%%%%%%

\begin{figure}[htb]
\includegraphics[width=0.9\columnwidth]{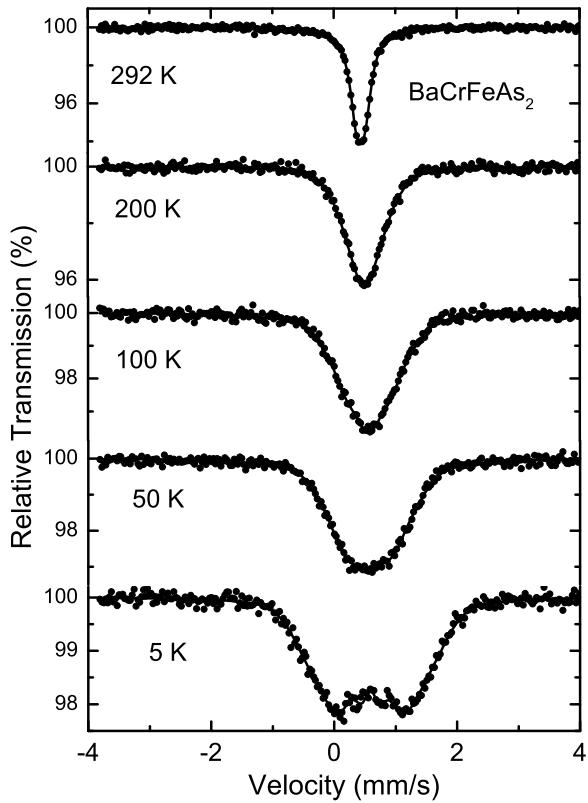}
\caption{M\"ossbauer spectra of BaCrFeAs$_2$ at the indicated temperatures.
Dots correspond to the experimental data, solid lines to the best fits.}
\label{fig:mosa} \end{figure}

\subsection{M\"ossbauer spectroscopy on BaCrFeAs$_2$}
\label{sec:mos}

Representative $^{57}$Fe M\"ossbauer spectra of BaCrFeAs$_2$ are shown in
Figure \ref{fig:mosa}. The room temperature spectrum appears as a broadened
line which can be fitted by a quadrupole doublet with an isomer shift
\textit{IS} of 0.44 mm\,s$^{-1}$ and a quadrupole splitting \textit{QS} of
0.13 mm\,s$^{-1}$. By contrast, the spectrum at 5\,K features a complex broad
pattern which evidences the presence of magnetic hyperfine splitting. The
spectrum was fitted by assuming a distribution of hyperfine fields, yielding
\textit{IS} = 0.59 mm\,s$^{-1}$ and a peak hyperfine field $B_\mathrm{hf} =
5.4$\,T. The isomer shifts are slightly larger than those of BaFe$_2$As$_2$
[\onlinecite{BLA11}] which adopts the SDW phase below the magnetic ordering
temperature. Nevertheless the local electronic structure is quite similar in
the two compounds. Most remarkably, the peak $B_\mathrm{hf}$ of about 5\,T in
BaCrFeAs$_2$ is nearly the same as the $B_\mathrm{hf}$ in the magnetically
ordered phases of BaFe$_2$As$_2$ [\onlinecite{BLA11}] and LaOFeAs
[\onlinecite{MCC08}]. From the neutron diffraction study we obtained an
average magnetic moment of 1.09 $\mu_\mathrm{B}$ per magnetic ion. Assuming
that the Cr moment is the same as in BaCr$_2$As$_2$ (1.8 $\mu_\mathrm{B}$) we
estimate the iron moment as 0.4 $\mu_\mathrm{B}$.

It has been pointed out that a direct estimation of Fe moments from
$B_\mathrm{hf}$ is questionable due to spin-orbit induced contributions to
$B_\mathrm{hf}$ in the iron pnictides [\onlinecite{Derondeau2016a}].
Nevertheless, a small iron moment is in qualitative agreement with the small
$B_\mathrm{hf}$ for BaCrFeAs$_2$. A similarly small iron moment of 0.4
$\mu_\mathrm{B}$ was reported for LaOFeAs [\onlinecite{CRU08}], whereas a
larger moment of 0.9 $\mu_\mathrm{B}$ was derived for BaFe$_2$As$_2$
[\onlinecite{HUA08}]. The spectra of BaCrFeAs$_2$ are, however, much less
resolved than those of BaFe$_2$As$_2$ and LaOFeAs which reflects the Fe-Cr
disorder in the present material giving rise to a broad $B_\mathrm{hf}$
distribution. Both, neutron diffraction and M\"ossbauer data confirm that Fe
and Cr atoms are disordered, while some previous electronic structure
calculations assumed an ordered arrangement [\onlinecite{SinghY2009a,HU10}]. A
large ordered Fe moment of 2.5--2.6 $\mu_\mathrm{B}$ predicted from the
electronic structure calculations for atomically ordered BaCrFeAs$_2$ (Ref.\
\onlinecite{HU10} and Table \ref{tab:BCFA}) is incompatible with the neutron
diffraction and the M\"ossbauer data.

\begin{figure}[htb]
\includegraphics[width=\columnwidth]{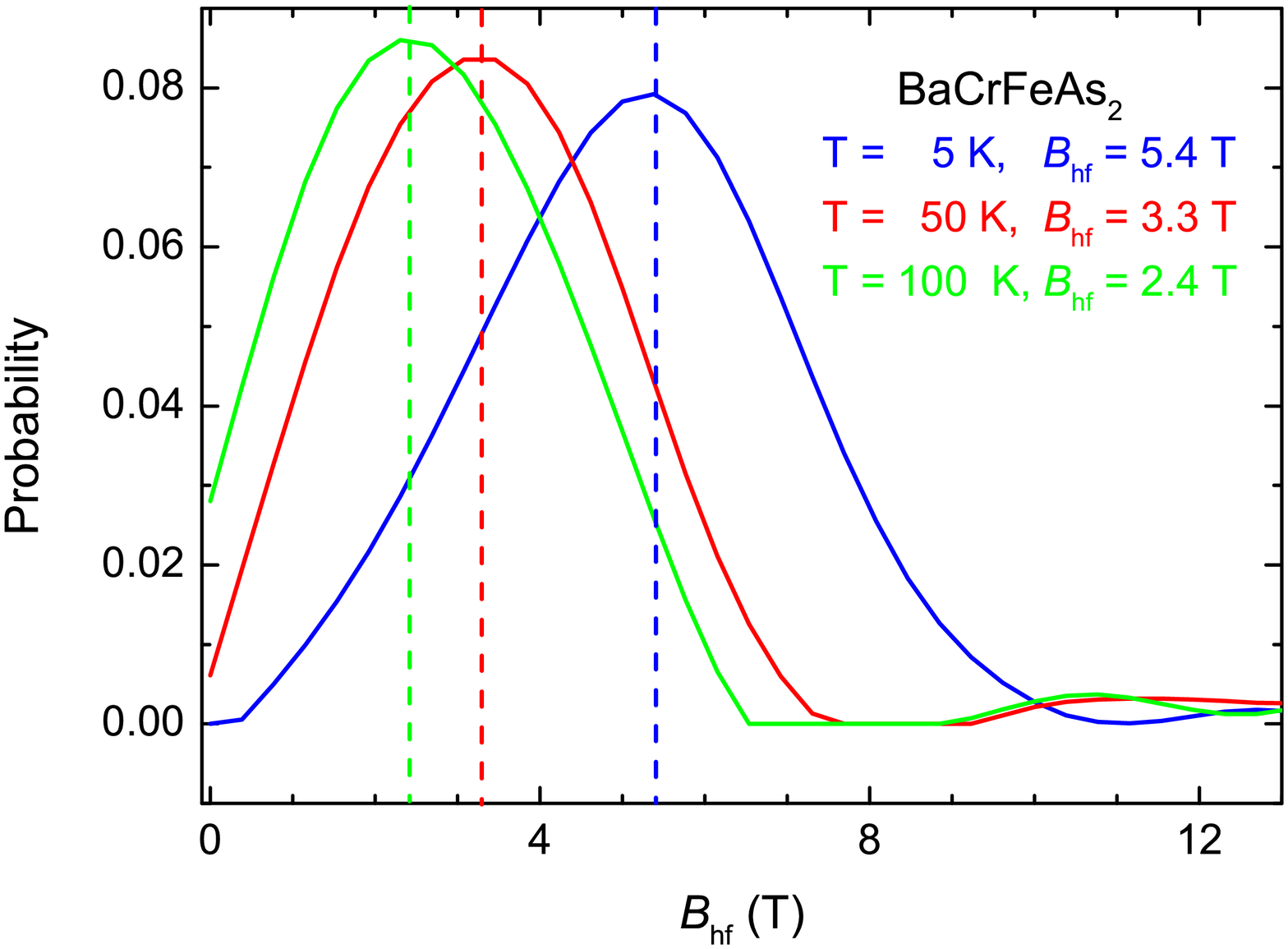}
\caption{Hyperfine field distributions in the low-temperature range which were
extracted from M\"ossbauer spectra of BaCrFeAs$_2$. The dashed lines indicate
the positions of the peak hyperfine fields given in the Figure.}
\label{fig:mosb} \end{figure}

Indications for an unusual behavior of the Fe moments is obtained from the
temperature dependence of $B_\mathrm{hf}$. With increasing temperature a
pronounced decrease in hyperfine splitting is apparent even in the temperature
region 5-100\,K (Figure \ref{fig:mosa}), which is unexpected as the N\'eel
temperature is much higher (265\,K). At 100\,K the $B_\mathrm{hf}$
distribution even extends down to 0\,T (Figure \ref{fig:mosb}) indicating that
a fraction of the iron atoms is already non-magnetic. These observations are
in contrast to the behavior of BaFe$_2$As$_2$ where in this temperature range
only a minor reduction of $B_\mathrm{hf}$ was observed although $T_\mathrm{N}$
is considerably smaller ($\approx 140$\,K) [\onlinecite{BLA11}]. The strong
decrease of the peak $B_\mathrm{hf}$ from 5.4\,T at 5\,K to 2.4\,T at 100\,K
is also in sharp contrast to the temperature dependence of the total magnetic
moments derived from neutron diffraction (Figure \ref{fig:Crmoment}).

A possible scenario is that the Fe moments do not participate in the $G$-type
afm structure and rather freeze in the low temperature range which is
consistent with the fact that no anomaly is apparent in the temperature
dependence of the moments between 2 and 100\,K. The formation of the $G$-type
afm order then is still driven by the large Cr moments. Further on, the upturn
in the susceptibility $\chi(T)$ below $T_\mathrm{N} = 265$\,K (Fig.\
\ref{fig:suscBCFA}), which is in contrast to the behavior of BaCr$_2$As$_2$,
is a further indication that the iron moments are at least partially
disordered. As demonstrated above electronic structure calculations of
atomically disordered BaCrFeAs$_2$ starting with zero Fe moment result in a
stable solution with $\mu_\mathrm{Fe} \sim 0$ and $\mu_\mathrm{Cr} =
2.76$\,$\mu_\mathrm{B}$. The resulting average moment of
1.38\,$\mu_\mathrm{B}$ per formula unit of BaCrFeAs$_2$ is not too far from
the experimental moment of 1.09\,$\mu_\mathrm{B}$. The mismatch of the Fe
moments with $G$-type antiferromagnetism explains why $T_\mathrm{N}$ in
Fe-substituted BaCr$_2$As$_2$ decreases drastically from 580\,K in the parent
compound to $\approx 50$\,K for an iron content of about 70\,\% where finally
the $G$-type order becomes unstable and is replaced by the SDW structure
[\onlinecite{Marty2011a}]. At higher temperature the hyperfine splitting
further decreases and for $T \geq 200$\,K essentially a broad unstructured
line is found. In this temperature range the Fe atoms are possibly polarized
by the Cr moments. Above 260\,K only minor changes in line broadening occur
which compares well with $T_\mathrm{N} = 265$\,K obtained from neutron
diffraction and susceptibility data.

%%%%%%%%%%%%%%%%%%%%%%%%%%%%%%%%%%%%%%%%%%%%%%%%%%%%%%%%%%%%%%%%%%%%%%%%%%%%%%%

\section{Conclusions}
\label{sec:conclusions}

We have studied in detail the structural, electronic, and magnetic properties
of BaCr$_2$As$_2$ and BaCrFeAs$_2$. BaCr$_2$As$_2$ as well as LaOCrAs
[\onlinecite{EDE16,PIZ16}] have been discussed recently in connection with the
Mott scenario (see Section \ref{sec:intro}) of transition-metal arsenide
superconductivity. Our powder neutron diffraction studies verify that
BaCr$_2$As$_2$ adopts the theoretically predicted $G$-type afm order at
$T_\mathrm{N} = 580$\,K and an ordered moment $\mu_\mathrm{Cr} =
1.9$\,$\mu_\mathrm{B}$ at 2\,K. Evidence for magneto-structural coupling
effects was found and it remains to be clarified whether this reflects some
type of electronic instability. The experimental magnetic Cr moment agrees
well with LDA electronic structure calculations whereas incorporation of
electron correlation within the LDA$+U$ scheme leads to Cr moments which are
too high. $T_\mathrm{N}$ is 1/3 lower than the ordering temperature calculated
within mean-field approximation, underlining the somewhat twodimensional
character of the magnetic system.

BaCrFeAs$_2$ still adopts this $G$-type afm structure but the ordering
temperature is less than half of that of BaCr$_2$As$_2$. The small average
moment $\mu_\mathrm{Cr/Fe} \approx 1.1$\,$\mu_\mathrm{B}$ in connection with
the small hyperfine field from M\"ossbauer spectra is in agreement with the
results of our calculations for atomically disordered BaCrFeAs$_2$ and an Fe
moment $\mu_\mathrm{Fe} \sim 0$. The N\'eel temperature calculated within
mean-field approximation is in good agreement with the experimental value
(260\,K and 265\,K, respectively). Also the average Cr/Fe moment is in fair
agreement with the calculation (2.76\,$\mu_\mathrm{B}$, distributed on two
atoms). These findings, together with an anomalous temperature dependence of
the hyperfine field, indicate that the small Fe moments are not incorporated
into the $G$-type afm order. Thus it may be conjectured that Fe favors the
stripe-type and Cr the checkerboard-type spin fluctuations, the latter being
considered to be detrimental to superconductivity
[\onlinecite{Fernandes2013a}]. In case of electron pairing mediated by
spin-fluctuations this finding would be in agreement with the fact that so far
no superconductivity was observed for BaCr$_2$As$_2$ or Cr-substituted
BaFe$_2$As$_2$. Similarly, no superconductivity was found in Mn-substituted
BaFe$_2$As$_2$, where the pure Mn compound BaMn$_2$As$_2$ adopts a stable
$G$-type afm order as well. Accordingly not only the doping level but also the
type of spin fluctuations has to be considered for predictions of new
superconducting compositions.

Both BaCr$_2$As$_2$ and BaCrFeAs$_2$ are metals. The linear specific heat
(Sommerfeld) coefficients $\gamma$ are much larger than expected from band
theory, by a factor of 2.4 for BaCr$_2$As$_2$ and a huge factor of 5.5 for
BaCrFeAs$_2$. The discrepancies for both compounds cannot be resolved by
reasonable strengths of correlations. However it seems that for BaCrFeAs$_2$
the disorder of the magnetic Cr and Fe atoms to some extent additionally
boosts the $\gamma$ value.

Our findings may be compared with results for the better investigated Mn 122
system. In contrast to BaCr$_2$As$_2$ BaMn$_2$As$_2$ is semiconducting
[\onlinecite{AnJ2009a,SinghY2009a,SinghY2009b,JohnstonDC2011a,ZhangWL2016a}].
The nominal $3d^5$ compound has a N\'eel temperature which is slightly higher
than that of BaCr$_2$As$_2$. It is possible to induce a metallic state in
BaMn$_2$As$_2$ by application of pressure [\onlinecite{Satya2011a}] or by
hole-doping with as little as 1.6\,\% of K [\onlinecite{Pandey2012a}]. In this
metallic state Ba$_{1-x}$K$_x$Mn$_2$As$_2$ is still $G$-type ordered
[\onlinecite{Lamsal2013a}] and is thus similar to BaCr$_2$As$_2$. But there is
one important difference, the Sommerfeld $\gamma$ in the metallic Mn system is
much smaller (8.4 mJ mol$^{-1}$ K$^{-1}$ for $x = 0.05$
[\onlinecite{Pandey2012a}]) than in BaCr$_2$As$_2$. This suggests that
electronic correlations are weak in the BaMn$_2$As$_2$ materials. A recent
photoemission (ARPES) study corroborates that there is almost no band
renormalization with respect to the DFT-based band structure in BaMn$_2$As$_2$
[\onlinecite{ZhangWL2016a}]. An ARPES study on BaCr$_2$As$_2$ would therefore
be highly welcome in order to shed more light on the development of electronic
correlations in the 122 family of transition-metal arsenides.

For Ba$_{1-x}$K$_x$Mn$_2$As$_2$ the N\'eel temperature and the ordered Mn
moment remain almost constant up to high substitution levels ($x \leq 0.4$)
[\onlinecite{Lamsal2013a}]. In contrast, in the substitution series
BaFe$_{2-x}$Mn$_x$As$_2$ for a low Mn substitution level $x = 0.15$ a clear
competition between stripe-type and checkerboard-type spin fluctuations could
be observed by inelastic neutron scattering [\onlinecite{Tucker2012a}], quite
similar to the BaCr$_{2-x}$Fe$_x$As$_2$ series [\onlinecite{Marty2011a}]. For
BaFe$_{2-x}$Mn$_x$As$_2$ a miscibility gap prevents the synthesis of
single-phase material for Mn contents $x > 0.24$ [\onlinecite{Pandey2011a}].

Although both modified BaMn$_2$As$_2$-based ($3d^5$) materials
[\onlinecite{Satya2011a,Pandey2012a,Lamsal2013a}] as well as BaCr$_2$As$_2$
are afm-ordered metals, until now there are no ways known to obtain
superconductors based on these materials. The partial substitution of Ba by K
in BaCr$_2$As$_2$ (hole doping) showed to be unsuccessful, but surprisingly
resulted in the discovery of superconducting K$_2$Cr$_3$As$_3$
[\onlinecite{BaoJK2015a,CaoGH2016a}]. Another way to possibly generate
superconductivity in Mn or Cr-based 122 or 1111-type arsenides is the
application of high pressure. With high pressure, in transition-metal
arsenides structural instabilities come into play. We have now undertaken a
structural and electrical transport study on one of our BaCr$_2$As$_2$
crystals under high pressure, the results of which will be reported elsewhere
[\onlinecite{MedvedevUnpubl}].

\acknowledgements

We thank Ralf Koban for assistence and Igor Veremchuk for spark plasma
sintering of a sample. MG acknowledges support of NSF-DMR 1507252 grant.

%%%%%%%%%%%%%%%%%%%%%%%%%%%%%%%%%%%%%%%%%%%%%%%%%%%%%%%%%%%%%%%%%%%%%%%%%%%%%%%

\end{document}